\documentclass[11pt]{article}

\usepackage[final]{acl}

\usepackage{times}
\usepackage{latexsym}
\usepackage[T1]{fontenc}
\usepackage[utf8]{inputenc}
\usepackage{microtype}
\usepackage{graphicx}

\usepackage{multirow}
\usepackage{enumitem}
\usepackage{xcolor}

\usepackage{amssymb}
\usepackage{amsmath}
\usepackage{booktabs}

\graphicspath{{./}}
\newcommand{\ignore}[1]{}

\AtBeginDocument{\definecolor{darkblue}{HTML}{000099}}

\title{Improving Long-Context Retrieval with Multi-Prefix Embedding}

\author{
  \textbf{Zhenglin Yu\textsuperscript{1}},
  \textbf{Xueguang Ma\textsuperscript{1}},
  \textbf{Shengyao Zhuang\textsuperscript{2}},
  \textbf{Zhichao Xu\textsuperscript{3}},
\\
  \textbf{Luyu Gao\textsuperscript{4}},
  \textbf{Crystina Zhang\textsuperscript{1}},
  \textbf{Jimmy Lin\textsuperscript{1}}
\\
\\
  \textsuperscript{1}University of Waterloo,
  \textsuperscript{2}University of Queensland,
  \textsuperscript{3}University of Utah,
  \textsuperscript{4}Carnegie Mellon University
\\
  \small{
    \textbf{Correspondence:} \href{mailto:x93ma@uwaterloo.ca}{x93ma@uwaterloo.ca}
  }
}

\begin{document}
\maketitle


\begin{abstract}
Long-context retrieval exposes a tension: single-vector embeddings lose fine-grained
detail, while token-level multi-vector methods incur prohibitive storage. We propose
Multi-Prefix Embedding (MPE), which partitions a document into chunks
separated by EOS tokens, encodes the full sequence in a single causal forward pass,
and extracts one embedding at each prefix boundary. MPE retains cross-chunk context,
enables chunk-level MaxSim matching, and trains with only document-level relevance
labels. Experiments on MLDR-en, BrowseComp-Plus, and LongEmbed show that MPE is
competitive with or outperforms single-vector, independent-chunk, and
multi-vector baselines, while providing a natural source attribution mechanism for locating evidence chunks.
\end{abstract}

\section{Introduction}
\label{sec:intro}

Long-context retrieval requires models to identify localized evidence within
documents that may span thousands of tokens. Dense retrieval encodes queries and
documents as fixed-dimensional vectors for efficient inner-product
search~\citep{karpukhin2020dense,izacard2022unsupervised,qwen3embedding,xu2025survey},
but compressing a long document into a single vector can obscure fine-grained
evidence~\citep{zhu2024longembed}. Token-level multi-vector methods such as
ColBERT~\citep{khattab2020colbert,santhanam-etal-2022-colbertv2} preserve finer
granularity, but incur substantially larger indexes and higher retrieval cost.
Chunk-based retrieval is cheaper, but independently encoded chunks lose cross-chunk
context and require post-hoc document-level aggregation. This raises a central
question: can we retain fine-grained matching while preserving context and keeping
the representation compact?

\begin{figure}[t]
  \centering
  \includegraphics[width=0.98\columnwidth]{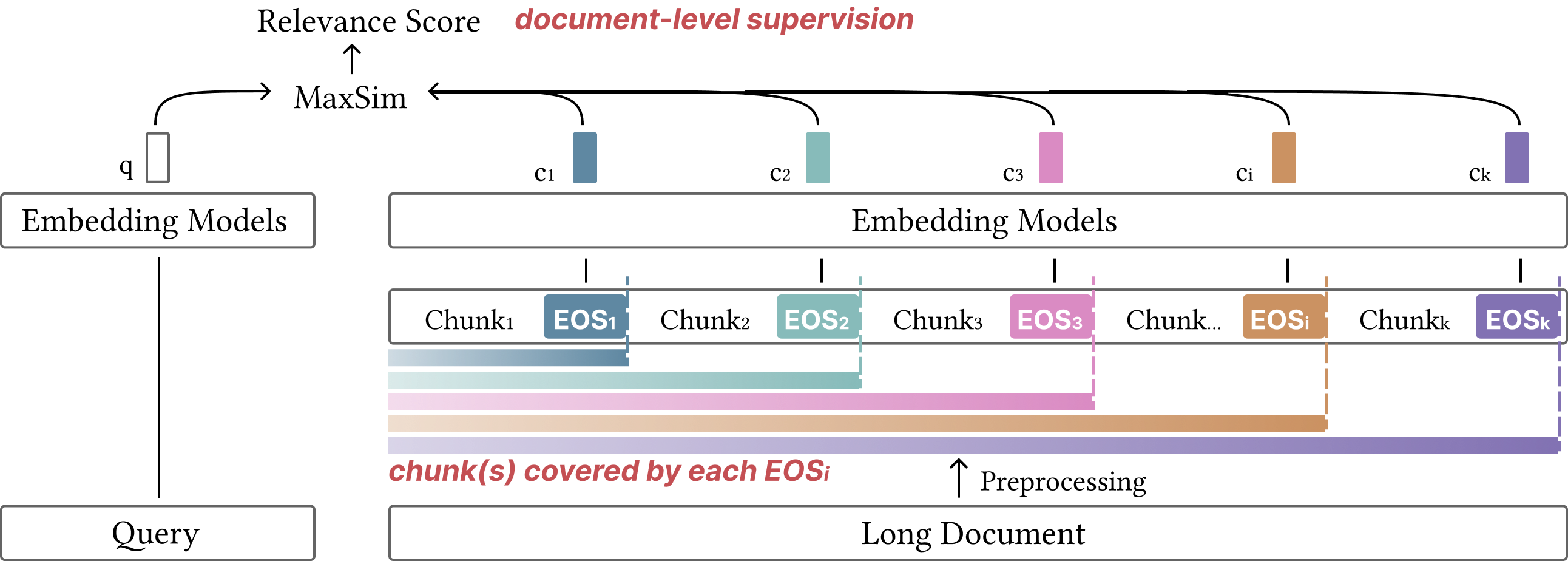}
  \caption{
Multi-Prefix Embedding (MPE). A long document is split into chunks separated by EOS
tokens and encoded in one causal forward pass. Prefix embeddings are extracted at
EOS$_1$, \ldots, EOS$_K$, and query--document relevance is computed via MaxSim.
  }
  \label{fig:mpe}
\end{figure}

We propose \textit{Multi-Prefix Embedding} (MPE), a compact multi-vector
representation for causal embedding models. As shown in Figure~\ref{fig:mpe}, MPE
splits a long document into chunks, inserts EOS tokens between them, and encodes the
resulting sequence in a single causal forward pass. The hidden state at each EOS
position summarizes the preceding content, yielding one prefix embedding per chunk
boundary without architectural changes or chunk-level annotations. Query--document
similarity is computed with MaxSim over the document's prefix embeddings, and the
model is trained using only document-level relevance labels.

MPE occupies a middle ground between single-vector, independent-chunk, and token-level
multi-vector retrieval. It provides multiple matching points within a long document,
preserves preceding cross-chunk context through causal attention, and stores only one
embedding per chunk boundary rather than one embedding per token. The MaxSim-selected
prefix also identifies the document region most responsible for a match, providing a
lightweight mechanism for surfacing candidate evidence.

We evaluate MPE on MLDR-en~\citep{chen2024bge},
BrowseComp-Plus~\citep{browsecompplus}, and
LongEmbed~\citep{zhu2024longembed}. Fine-tuning Qwen3-Embedding-0.6B~\citep{qwen3embedding}
on MLDR-en, MPE matches or exceeds single-vector and independent-chunk baselines on
most benchmarks, while random prefix-length augmentation improves robustness under
granularity mismatch, especially on out-of-domain BrowseComp-Plus.

Our main contributions are:
\begin{enumerate}[leftmargin=*,nosep]
\item We propose MPE, a compact multi-vector representation for long-context
  retrieval that balances fine-grained matching, context preservation, and indexing
  efficiency.
\item We show that MaxSim-selected prefixes often align with annotated answer
  locations, suggesting a lightweight mechanism for candidate evidence surfacing.
\item We present experiments and ablations across representation methods, attention
  mechanisms, and granularity settings.
\end{enumerate}

\section{Related Work}
\label{sec:related}

Dense retrieval encodes queries and documents into fixed-dimensional vectors for
inner-product search~\citep{karpukhin2020dense,izacard2022unsupervised}. DPR uses
BERT encoders with [CLS] pooling~\citep{karpukhin2020dense}, while recent
decoder-only embedding models such as RepLLaMA~\citep{repllama},
E5-Mistral~\citep{wang2024improving}, and Qwen3-Embedding~\citep{qwen3embedding}
use last-token pooling. Although efficient, single-vector embeddings can obscure
fine-grained evidence in long documents.

Multi-vector methods improve fine-grained matching by representing each document
with multiple embeddings. ColBERT~\citep{khattab2020colbert,santhanam-etal-2022-colbertv2}
uses late interaction over token-level representations, but its storage and
retrieval cost grows with document length. MaxP~\citep{dai2019deepertextunderstanding}
splits documents into fixed-size chunks and takes the maximum independently encoded
chunk score, but loses cross-chunk context.

MPE is most related to compact and contextual multi-vector representations. BGE
Landmark Embedding~\citep{luo2024bge} also extracts hidden states from special
positions in a causal LLM, but targets context selection for retrieval-augmented
long-context LLMs using landmark tokens and multi-stage supervision. In contrast,
MPE uses EOS-separated chunk boundaries for document-level retrieval and trains with
a single contrastive objective over document-level labels. Late
chunking~\citep{gunther2024late} pools chunk representations after bidirectional
document encoding, while token
pooling~\citep{clavie2024reducingfootprintmultivectorretrieval} compresses
ColBERT-style token embeddings for short-passage retrieval.


\section{Method}
\label{sec:method}

MPE consists of EOS-based prefix construction, MaxSim scoring, and random
prefix-length augmentation. Figure~\ref{fig:mpe}
illustrates the overall framework.

Given a document $p=[t_1,\ldots,t_L]$, we split it into $K$ consecutive chunks of
$m$ content tokens and append an EOS token after each chunk:
\begin{align}
\tiny
  x = [\,t_1\ldots t_m,\texttt{<eos>},\,
        t_{m+1}\ldots t_{2m},\texttt{<eos>}\ldots\,]
\end{align}
The resulting sequence is encoded in a single forward pass through a pretrained
causal language model. Under causal masking, the hidden state at each EOS position
attends to the current chunk and all preceding chunks, yielding a contextualized
prefix representation. Unlike independently encoded chunks, each prefix embedding
therefore incorporates preceding cross-chunk context. We L2-normalize the hidden
states at EOS positions to obtain document prefix embeddings
$\{c_1,\ldots,c_K\}$, where $c_k\in\mathbb{R}^d$. When $K=1$, MPE reduces to
standard last-token pooling.

\subsection{Scoring, Training, and Retrieval}
\label{sec:maxsim}

Given a query embedding $q$, MPE scores a document by MaxSim over its prefix
embeddings:
\begin{align}
\tiny
  s(q,p)=\operatorname{MaxSim}(q,p)
  =\max_{k\in\{1,\ldots,K\}} q^\top c_k 
\end{align}
This allows a query to match the most relevant document region rather than requiring
the whole document to be compressed into one vector. We train MPE with contrastive
loss over document-level relevance labels using cross-device in-batch negatives.

At retrieval time, all prefix embeddings are indexed in a
FAISS~\citep{johnson2019billion} flat inner-product index together with their
document IDs. Prefix-level hits are aggregated by document ID using the maximum
prefix score to produce the final document ranking.

\subsection{Random Prefix-Length Augmentation}
\label{sec:rand-aug}

Fixed chunk sizes can make the model sensitive to a specific granularity. To improve
robustness, we sample the content chunk length for each training passage,
$m\sim\mathcal{U}[m_{\min},m_{\max}]$, and construct EOS-separated prefixes using
that length. All chunks within a passage share the same sampled $m$. This exposes
the model to diverse prefix boundaries during training and helps a single MPE model
generalize across inference chunk sizes without retraining.

\section{Experiments}
\label{sec:experiments}

\subsection{Setup}
\label{sec:setup}

\textbf{Datasets.}
We evaluate on three long-context retrieval benchmarks. \textit{MLDR-en}~\citep{chen2024bge}
contains long-document retrieval queries with documents up to 8{,}192 tokens; we use
its 800-query test split. \textit{BrowseComp-Plus}~\citep{browsecompplus} contains
830 difficult web queries with human-verified supporting documents and mined hard
negatives. \textit{LongEmbed}~\citep{zhu2024longembed} evaluates long-context
retrieval ability; we report results on the real-document subsets NarrativeQA,
2WikiMQA, SummScreen, and QMSum.

\textbf{Model and training.}
We use Qwen3-Embedding-0.6B~\citep{qwen3embedding} as the base model and fine-tune it
on the MLDR-en training split for one epoch using Tevatron~\citep{tevatron}. Training
uses LoRA adapters with rank 16, $\alpha=64$, and dropout 0.1 on all attention and
MLP projections (\texttt{q/k/v/o/gate/up/down\_proj}). We use learning rate
$1{\times}10^{-4}$, temperature $\tau=0.03$, per-GPU batch size 2, group size 4,
query length 512, maximum document length 8{,}192, right-side padding, and gradient
checkpointing. Experiments were run on 8 NVIDIA RTX 5090 GPUs. The total compute budget for the reported experiments is approximately 120 GPU-hours, including training, ablations, and inference.

For zero-shot evaluation, we reuse the MLDR-en fine-tuned model without further
training. We use maximum document length 4{,}096 for BrowseComp-Plus and 8{,}192 for
LongEmbed.

\textbf{Configurations.}
We compare five configurations that isolate the effects of multi-vector inference,
MaxSim training, cross-chunk attention, and random prefix-length augmentation
(Table~\ref{tab:configs}).

\begin{table}[ht]
\centering
\setlength{\tabcolsep}{3pt}
\renewcommand{\arraystretch}{0.95}
\resizebox{\columnwidth}{!}{%
\begin{tabular}{@{}lccc@{}}
\toprule
Setting & MaxSim Train & Cross-Chunk Attn. & Rand. Size \\
\midrule
Single-vector   & \texttimes & -- & \texttimes \\
MaxP            & \texttimes & \texttimes & \texttimes \\
MaxP-Train      & \checkmark & \texttimes & \texttimes \\
MPE Fixed-$N$   & \checkmark & \checkmark & \texttimes \\
MPE-Rand$[a,b]$ & \checkmark & \checkmark & \checkmark \\
\bottomrule
\end{tabular}%
}
\caption{Comparison of configuration settings.}
\label{tab:configs}
\end{table}

\noindent
\textit{Single-vector} uses last-token pooling without chunking.
\textit{MaxP}~\citep{dai2019deepertextunderstanding} encodes chunks independently at
inference and takes the maximum chunk score. \textit{MaxP-Train} adds MaxSim
training. \textit{MPE Fixed-$N$} trains and infers with fixed EOS-separated prefixes,
while \textit{MPE-Rand$[a,b]$} samples the training chunk size from $[a,b]$. Unless
specified, chunk-based and MPE methods use inference chunk size 64; LongEmbed and
BrowseComp-Plus are zero-shot after MLDR-en training.
\subsection{Main Results}
\label{sec:main-results}

\begin{table*}[t]
\centering
\setlength{\tabcolsep}{4pt}
\renewcommand{\arraystretch}{1}
\resizebox{\textwidth}{!}{%
\begin{tabular}{lccccccc}
\toprule
\textbf{Method} & \textbf{MLDR Train} & \textbf{MLDR-en} & \textbf{BrowseComp-Plus} & \textbf{NarrativeQA} & \textbf{2WikiMQA} & \textbf{SummScreen} & \textbf{QMSum}\\
\midrule
BM25                                                  & $\times$    & 0.679 & 0.016 & 0.715          & 0.965          & 0.976          & 0.813 \\
jina-v2-base-en~\citep{gunther2023jina}              & $\times$    & 0.370 & --    & 0.394          & 0.740          & 0.935          & 0.389 \\
OpenAI-Ada-002                                       & --          & 0.387 & --    & 0.411          & 0.801          & 0.918          & 0.400 \\
E5-Mistral-7B~\citep{wang2024improving}              & $\times$    & 0.433 & --    & 0.449          & --             & --             & --    \\
BGE-M3~\citep{chen2024bge}                           & \checkmark  & 0.489 & --    & 0.487          & 0.780          & 0.940          & 0.355 \\
BGE-M3$_\text{mv}$~\citep{chen2024bge}               & \checkmark  & 0.558 & --    & 0.554          & --             & --             & --    \\
\midrule
Single-vector                                        & \checkmark  & 0.548 & 0.132 & 0.543          & 0.863          & \textbf{0.981} & 0.565 \\
MaxP                                                 & \checkmark  & 0.758 & 0.103 & \textbf{0.604} & 0.944          & 0.906          & 0.550 \\
MaxP-Train                                           & \checkmark  & 0.776 & 0.104 & 0.557          & 0.945          & 0.903          & 0.570 \\
MPE Fixed-64                                         & \checkmark  & \textbf{0.783} & 0.122 & 0.580 & \textbf{0.948} & 0.928          & 0.652 \\
MPE-Rand$[32,1024]$                                  & \checkmark  & 0.760 & \textbf{0.153} & 0.579 & 0.945          & 0.949          & \textbf{0.705} \\
\bottomrule
\end{tabular}%
}
\caption{
nDCG@10 on long-context retrieval benchmarks.
Baseline results are from original papers~\citep{gunther2023jina,zhu2024longembed,chen2024bge,wang2024improving,nussbaum2024nomic}.
BGE-M3$_\text{mv}$ denotes ColBERT-style multi-vector retrieval.
Methods below the midrule use Qwen3-Embedding-0.6B; chunk-based and MPE methods use
inference chunk size 64 unless otherwise specified. Bold marks the best dense result.
}
\label{tab:mpe-main-idea}
\end{table*}

Table~\ref{tab:mpe-main-idea} reports nDCG@10 on in-domain and zero-shot
long-context retrieval benchmarks. MPE Fixed-64 achieves the best dense result on
MLDR-en, while MPE-Rand$[32,1024]$ is the only multi-vector variant to outperform
Single-vector on BrowseComp-Plus, highlighting the importance of granularity
robustness for out-of-domain retrieval. On LongEmbed, MPE gives the best
dense result on 2WikiMQA and QMSum and remains competitive on NarrativeQA; SummScreen
is nearly saturated by BM25 and Single-vector retrieval.

MPE improves over independent chunking through contextual prefix embeddings and
granularity robustness. Under the same inference chunk size, MPE Fixed-64 outperforms
MaxP-Train on MLDR-en, suggesting that preceding cross-chunk context helps.
MPE-Rand sacrifices some in-domain peak performance on MLDR-en but improves
BrowseComp-Plus and QMSum.

\begin{figure}[t]
  \centering
  \includegraphics[width=0.95\columnwidth]{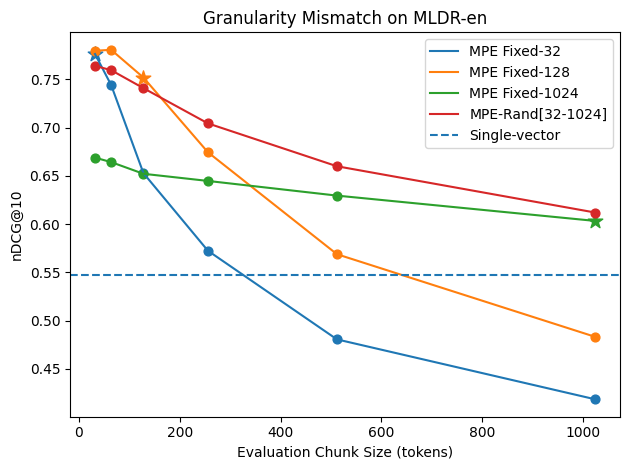}
  \caption{
  Granularity mismatch on MLDR-en. Fixed-size MPE degrades under mismatched
  granularities, while MPE-Rand tracks the upper envelope with a single model.
  \textit{Star} symbols denote matched train--eval sizes; \textit{circle} symbols
  denote mismatched sizes.
  }
  \label{fig:granularity_mismatch}
\end{figure}

\paragraph{Granularity robustness.}
Figure~\ref{fig:granularity_mismatch} varies the inference chunk size from 32 to
1024 on MLDR-en. Fixed-size MPE models peak near their training granularity but
degrade under mismatch, while MPE-Rand$[32,1024]$ tracks their upper envelope with a
single checkpoint.

\paragraph{Causal vs.\ bidirectional attention.}
As a diagnostic, we replace the causal self-attention mask with a bidirectional mask
during fine-tuning and inference. This modification reduces MLDR-en nDCG@10 from
0.783 to 0.694, suggesting that MPE benefits from the pretrained causal attention
structure rather than simply from exposing each EOS position to more tokens.

\paragraph{Storage overhead.}
MPE stores $K=T/C$ vectors per document, where $T$ is document length and $C$ is
chunk size. On MLDR-en ($T\leq8{,}192$), this gives $K\approx128$ at $C=64$ and
$K\approx16$ at $C=512$. For $N$ documents, MPE stores $\mathcal{O}(KN)$ vectors,
between single-vector retrieval $\mathcal{O}(N)$ and token-level late interaction
$\mathcal{O}(TN)$, with $K\ll T$.

\begin{figure}[t]
  \centering
  \includegraphics[width=0.95\columnwidth]{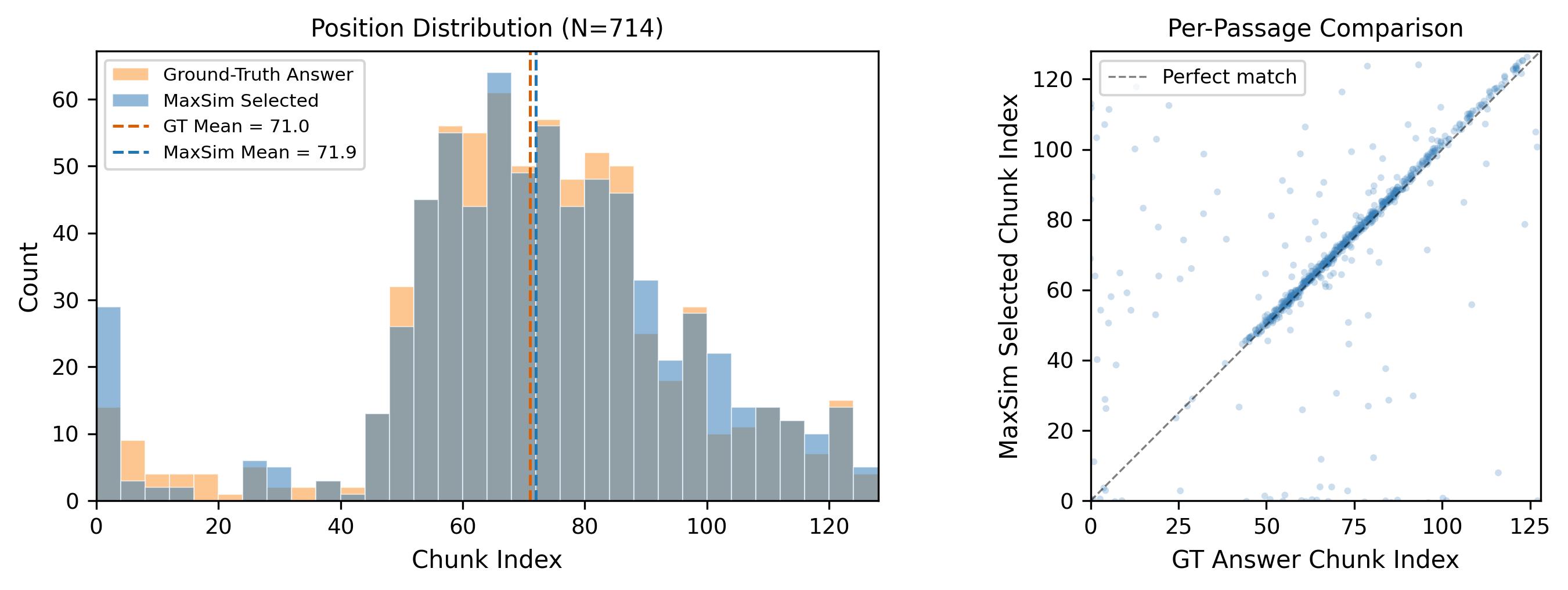}
  \caption{
  MaxSim-selected chunk positions vs.\ Gemini-annotated answer positions on MLDR-en.
  Left: overlaid position distributions. Right: per-passage scatter plot showing
  substantial rank correlation (Spearman $\rho=0.77$).
  }
  \label{fig:maxsim-vs-gt-chunk}
\end{figure}

\paragraph{Source attribution.}
Using Gemini-annotated answer spans on 714 MLDR-en passages, MaxSim selects a chunk
within $\pm1$ position of the annotated chunk for 65.7\% of passages, with Spearman
correlation $\rho=0.77$ (Figure~\ref{fig:maxsim-vs-gt-chunk}). Although the annotations are LLM-derived, this suggests that MPE can surface candidate evidence in addition to document-level rankings.

\paragraph{End-to-end search agent.}
Table~\ref{tab:browsecomp-e2e} evaluates MPE-Rand in a BrowseComp-Plus search-agent
setting. Replacing the single-vector retriever with MPE-Rand improves answer
accuracy from 42.29\% to 51.45\%, supporting-document recall from 48.00\% to
57.70\%, and reduces retrieval calls.

\begin{table}[t]
  \centering
  \setlength{\tabcolsep}{3pt}
  \renewcommand{\arraystretch}{0.95}
  \resizebox{\columnwidth}{!}{%
  \begin{tabular}{lcccc}
    \toprule
    \textbf{Retriever} & \textbf{Acc.\,(\%)} & \textbf{Recall\,(\%)} & \textbf{Avg.\ Calls} & \textbf{Cal.\ Err.\,(\%)}\footnotemark \\
    \midrule
    BM25          & 34.94 & 37.78 & 18.85 & 51.17 \\
    Single-vector & 42.29 & 48.00 & 17.40 & 46.99 \\
    MPE-Rand      & \textbf{51.45} & \textbf{57.70} & \textbf{15.78} & \textbf{39.74} \\
    \bottomrule
  \end{tabular}%
  }
  \caption{
  End-to-end BrowseComp-Plus with a Gemini 3 Flash Preview search agent. Accuracy is
  judged by GPT-4o-mini. MPE-Rand is trained on $[32,1024]$ and evaluated at chunk
  size 64.
  }
  \label{tab:browsecomp-e2e}
\end{table}
\footnotetext{
Calibration error measures how closely predicted confidence matches actual answer accuracy \citep{browsecompplus}.
}

\section{Conclusion}

We presented Multi-Prefix Embedding (MPE), a compact chunk-level multi-vector method
for long-document retrieval. By inserting EOS tokens between chunks, MPE extracts
prefix embeddings in a single causal LM forward pass, preserving preceding
cross-chunk context while enabling fine-grained MaxSim matching without chunk-level
supervision or architectural changes.

Experiments on MLDR-en, BrowseComp-Plus, and LongEmbed show that MPE matches or
outperforms single-vector and independent-chunk baselines on most benchmarks. Random
prefix-length augmentation improves robustness across chunk sizes, with particularly
strong gains under granularity mismatch and out-of-domain retrieval. MaxSim-selected
prefixes align with Gemini annotated answer locations.

Code and model checkpoints will be released upon publication.

\newpage
\section*{Limitations}

Our experiments are limited to a single 0.6B-parameter causal embedding model and
English-language benchmarks. Scaling MPE to larger embedding models, multilingual
retrieval, and stronger instruction-tuned retrievers remains future work.

MPE introduces additional storage and retrieval cost compared with single-vector
embeddings because each document is represented by multiple prefix embeddings. While
this cost is substantially smaller than token-level late interaction, practical
deployment may require product quantization, residual compression, pruning, or
adaptive prefix selection.

Our comparisons against Landmark Embedding~\citep{luo2024bge} and Late
Chunking~\citep{gunther2024late} are based on reported results and conceptual
differences rather than a fully controlled head-to-head evaluation. A matched
comparison under the same backbone, training data, context length, and indexing
budget would better isolate the effect of MPE's prefix representation.

Finally, our source-attribution analysis relies on Gemini-generated answer-span
annotations. Although the observed correlation suggests that MaxSim often selects
meaningful evidence locations, these labels are not human gold annotations and may
contain noise.

\section*{Ethics Descriptions}
This work uses public retrieval benchmarks and does not involve human-subject data collection. Like other retrieval systems, MPE may amplify biases or misinformation present in training corpora or retrieved documents, especially when used in downstream search or question-answering agents. Our BrowseComp-Plus agent evaluation uses LLM-based answer judging, which can introduce noise or evaluator bias; we therefore treat it as complementary to retrieval metrics rather than definitive human evaluation. We disclose the main compute setup in Section~\ref{sec:setup}; practical deployments should consider the storage, latency, and energy costs of multi-vector indexing.

\bibliography{references}

@article{zhu2024longembed,
  title={LongEmbed: Extending Embedding Models for Long Context Retrieval},
  author={Zhu, Dawei and Wang, Liang and Yang, Nan and Song, Yifan and Wu, Wenhao and Wei, Furu and Li, Sujian},
  journal={arXiv preprint arXiv:2404.12096},
  year={2024},
  url={https://arxiv.org/abs/2404.12096}
}

@inproceedings{repllama,
author = {Ma, Xueguang and Wang, Liang and Yang, Nan and Wei, Furu and Lin, Jimmy},
title = {Fine-Tuning LLaMA for Multi-Stage Text Retrieval},
year = {2024},
isbn = {9798400704314},
publisher = {Association for Computing Machinery},
address = {New York, NY, USA},
url = {https://doi.org/10.1145/3626772.3657951},
doi = {10.1145/3626772.3657951},
booktitle = {Proceedings of the 47th International ACM SIGIR Conference on Research and Development in Information Retrieval},
pages = {2421–2425},
numpages = {5},
location = {Washington DC, USA},
series = {SIGIR '24}
}

@article{clavie2024reducingfootprintmultivectorretrieval,
      title={Reducing the Footprint of Multi-Vector Retrieval with Minimal Performance Impact via Token Pooling}, 
      author={Benjamin Clavié and Antoine Chaffin and Griffin Adams},
      year={2024},
      journal={arXiv:2409.14683},
      url={https://arxiv.org/abs/2409.14683}, 
}

@inproceedings{karpukhin2020dense,
  title={Dense Passage Retrieval for Open-Domain Question Answering},
  author={Karpukhin, Vladimir and O{\u{g}}uz, Barlas and Min, Sewon and Lewis, Patrick and Wu, Ledell and Edunov, Sergey and Chen, Danqi and Yih, Wen-tau},
  booktitle={Proceedings of the 2020 Conference on Empirical Methods in Natural Language Processing (EMNLP)},
  pages={6769--6781},
  year={2020},
  url={https://aclanthology.org/2020.emnlp-main.550/},
  doi={10.18653/v1/2020.emnlp-main.550}
}

@article{izacard2022unsupervised,
  title={Unsupervised Dense Information Retrieval with Contrastive Learning},
  author={Izacard, Gautier and Caron, Mathilde and Hosseini, Lucas and Riedel, Sebastian and Bojanowski, Piotr and Joulin, Armand and Grave, Edouard},
  journal={Transactions on Machine Learning Research},
  year={2022},
  url={https://arxiv.org/abs/2112.09118}
}

@inproceedings{khattab2020colbert,
  title={{ColBERT}: Efficient and Effective Passage Search via Contextualized Late Interaction over {BERT}},
  author={Khattab, Omar and Zaharia, Matei},
  booktitle={Proceedings of the 43rd International ACM SIGIR Conference on Research and Development in Information Retrieval},
  pages={39--48},
  year={2020},
  doi={10.1145/3397271.3401075}
}

@article{gunther2023jina,
  title={Jina Embeddings 2: 8192-Token General-Purpose Text Embeddings for Long Documents},
  author={G{\"u}nther, Michael and Ong, Jackmin and Mohr, Isabelle and Abdessalem, Alaeddine and Abel, Tanguy and Akber, Mohammad Afzal and Gusber, Susana and Mastrapas, Georgios and Milliken, Robin and Wang, Bo},
  journal={arXiv preprint arXiv:2310.19923},
  year={2023},
  url={https://arxiv.org/abs/2310.19923}
}

@article{gunther2024late,
  title={Late Chunking: Contextual Chunk Embeddings Using Long-Context Embedding Models},
  author={G{\"u}nther, Michael and Mohr, Isabelle and Williams, Daniel J and Wang, Bo and Rodr{\'\i}guez, Han Xiao},
  journal={arXiv preprint arXiv:2409.04701},
  year={2024},
  url={https://arxiv.org/abs/2409.04701}
}

@article{luo2024bge,
  title={{BGE} Landmark Embedding: A Chunking-Free Embedding Method For Retrieval Augmented Long-Context Large Language Models},
  author={Luo, Kun and Liu, Zheng and Xiao, Shitao and Liu, Kang},
  journal={arXiv preprint arXiv:2402.11573},
  year={2024},
  url={https://arxiv.org/abs/2402.11573}
}

@article{qwen3embedding,
  title={Qwen3 Embedding: Advancing Text Embedding and Reranking Through Foundation Models},
  author={{Qwen Team}},
  journal={arXiv preprint arXiv:2506.05176},
  year={2025},
  url={https://arxiv.org/abs/2506.05176}
}

@article{chen2024bge,
  title={{BGE} {M3}-Embedding: Multi-Lingual, Multi-Functionality, Multi-Granularity Text Embeddings Through Self-Knowledge Distillation},
  author={Chen, Jianlv and Xiao, Shitao and Zhang, Peitian and Luo, Kun and Lian, Defu and Liu, Zheng},
  journal={arXiv preprint arXiv:2402.03216},
  year={2024},
  url={https://arxiv.org/abs/2402.03216}
}

@article{browsecompplus,
  title={{BrowseComp-Plus}: A More Fair and Transparent Evaluation Benchmark of Deep-Research Agent},
  author={Chen, Zijian and Ma, Xueguang and Zhuang, Shengyao and Nie, Ping and Zou, Kai and Liu, Andrew and Green, Joshua and Patel, Kshama and Meng, Ruoxi and Su, Mingyi and Sharifymoghaddam, Sahel and Li, Yanxi and Hong, Haoran and Shi, Xinyu and Liu, Xuye and Thakur, Nandan and Zhang, Crystina and Gao, Luyu and Chen, Wenhu and Lin, Jimmy},
  journal={arXiv preprint arXiv:2508.06600},
  year={2025},
  url={https://arxiv.org/abs/2508.06600}
}

@article{johnson2019billion,
  title={Billion-scale similarity search with {GPUs}},
  author={Johnson, Jeff and Douze, Matthijs and J{\'e}gou, Herv{\'e}},
  journal={IEEE Transactions on Big Data},
  volume={7},
  number={3},
  pages={535--547},
  year={2019},
  publisher={IEEE},
  doi={10.1109/TBDATA.2019.2921572}
}

@article{nussbaum2024nomic,
  title={Nomic Embed: Training a Reproducible Long Context Text Embedder},
  author={Nussbaum, Zach and Morris, John X and Duderstadt, Brandon and Mulyar, Andriy},
  journal={arXiv preprint arXiv:2402.01613},
  year={2024},
  url={https://arxiv.org/abs/2402.01613}
}

@article{wang2024improving,
  title={Improving Text Embeddings with Large Language Models},
  author={Wang, Liang and Yang, Nan and Huang, Xiaolong and Yang, Linjun and Majumder, Rangan and Wei, Furu},
  journal={arXiv preprint arXiv:2401.00368},
  year={2024},
  url={https://arxiv.org/abs/2401.00368}
}

@article{tevatron,
  title={Tevatron 2.0: Unified Document Retrieval Toolkit across Scale, Language, and Modality},
  author={Ma, Xueguang and Gao, Luyu and Zhuang, Shengyao and Zhan, Jiaqi Samantha and Callan, Jamie and Lin, Jimmy},
  journal={arXiv preprint arXiv:2505.02466},
  year={2025},
  url={https://arxiv.org/abs/2505.02466}
}

@inproceedings{dai2019deepertextunderstanding,
author = {Dai, Zhuyun and Callan, Jamie},
title = {Deeper Text Understanding for IR with Contextual Neural Language Modeling},
year = {2019},
isbn = {9781450361729},
publisher = {Association for Computing Machinery},
address = {New York, NY, USA},
url = {https://doi.org/10.1145/3331184.3331303},
doi = {10.1145/3331184.3331303},
booktitle = {Proceedings of the 42nd International ACM SIGIR Conference on Research and Development in Information Retrieval},
pages = {985–988},
numpages = {4},
location = {Paris, France},
series = {SIGIR'19}
}

@inproceedings{santhanam-etal-2022-colbertv2,
    title = "{C}ol{BERT}v2: Effective and Efficient Retrieval via Lightweight Late Interaction",
    author = "Santhanam, Keshav  and
      Khattab, Omar  and
      Saad-Falcon, Jon  and
      Potts, Christopher  and
      Zaharia, Matei",
    editor = "Carpuat, Marine  and
      de Marneffe, Marie-Catherine  and
      Meza Ruiz, Ivan Vladimir",
    booktitle = "Proceedings of the 2022 Conference of the North American Chapter of the Association for Computational Linguistics: Human Language Technologies",
    month = jul,
    year = "2022",
    address = "Seattle, United States",
    publisher = "Association for Computational Linguistics",
    url = "https://aclanthology.org/2022.naacl-main.272/",
    doi = "10.18653/v1/2022.naacl-main.272",
    pages = "3715--3734"
}

@article{xu2025survey,
    title={A Survey of Model Architectures in Information Retrieval},
    author={Xu, Zhichao and Mo, Fengran and Huang, Zhiqi and Zhang, Crystina and Yu, Puxuan and Wang, Bei and Lin, Jimmy and Srikumar, Vivek},
    journal={arXiv preprint arXiv:2502.14822},
    year={2025},
    url={https://arxiv.org/abs/2502.14822}
}


\appendix

\section{License}
\label{app:license}

We use the following external artifacts: MLDR-en, BrowseComp-Plus, LongEmbed, and
Qwen3-Embedding-0.6B. MLDR-en and BrowseComp-Plus are released under the MIT
license. Qwen3-Embedding-0.6B is released under the Apache-2.0 license. LongEmbed
is a public research benchmark for long-context retrieval evaluation; its Hugging
Face dataset card does not list an explicit license field, so we use it only for
research evaluation and follow the terms and licenses of the original component
datasets included in LongEmbed.

Our use of these artifacts is consistent with their intended research purpose:
datasets are used for retrieval benchmarking and evaluation, and the pretrained
Qwen3-Embedding model is used as a base embedding model for research experiments.
We do not redistribute the original benchmark data beyond what is permitted by the
artifact providers. Any released code or checkpoints from this work are intended
for research and reproducibility, and users should comply with the licenses and
terms of the underlying datasets, pretrained models, and software dependencies.

\section{LLM Usage}
\label{app:llm-usage}

The authors used generative AI tools for language polishing, grammar checking,
minor editing suggestions, and assistance with code debugging. The tools were not
used to generate the core technical ideas, design the experiments, produce the
reported results, or make final scientific claims. All experiments, analyses,
results, and final writing decisions were performed and verified by the authors.

\end{document}